\documentclass[pra,twocolumn,groupedaddress,showpacs,floatfix]{revtex4}

\usepackage{graphics}
\usepackage{graphicx}
\usepackage{bm}
\usepackage{amsmath}
\usepackage{amsfonts}
\usepackage{amssymb}
\usepackage{latexsym}

\begin{document}

\title{Speeding up gate operations through dissipation\footnote{It is a pleasure to contribute this article to this special issue celebrating Herbert Walther's birthday: Professor Walther's group have pioneered research in cavity QED and in the study of laser-coded trapped ions, and his vision of how these two fields can be synthesized and combined have stimulated our own interest in this field.}}
\author{Almut Beige$^{1,2}$, Hugo Cable$^1$, Carsten Marr$^3$\footnote{Present address: Bioinformatics group, Darmstadt University of Technology, D-64287 Darmstadt, Germany}, and Peter L. Knight$^1$}
\affiliation{$^1$Blackett Laboratory, Imperial College London, Prince Consort Road, London SW7 2BW, United Kingdom\\
$^2$Department of Applied Mathematics and Theoretical Physics, University of Cambridge, Wilberforce Road, Cambridge CB3 0WA, United Kingdom\\
$^3$Max-Planck-Institut f\"ur Quantenoptik, Hans-Kopfermann-Str. 1, 85748 Garching, Germany} 

\date{\today}

\begin{abstract}
It is commonly believed that decoherence is the main obstacle to quantum information processing. In contrast to this, we show how decoherence in the form of dissipation can improve the performance of certain quantum gates. As an example we consider the realisations of a controlled phase gate and a two-qubit SWAP operation with the help of a single laser pulse in atom-cavity systems. In the presence of spontaneous decay rates, the speed of the gates can be improved by a factor 2 without sacrificing high fidelity and robustness against parameter fluctuations. Even though this leads to finite gate failure rates, the scheme is comparable with other quantum computing proposals. 
\end{abstract}

\pacs{42.50.Lc, 03.67.Mn, 03.67.Pp}

\maketitle

\section{Introduction}

Quantum computing is a paradigm in which quantum entanglement and interference are exploited for information processing. Algorithms have been proposed which can even be exponentially more effective than the best known classical solutions \cite{shor,deutsch,grover}. Elements of quantum computing have been demonstrated experimentally using nuclear magnetic resonance \cite{Chuang} and trapped ions \cite{exps,exps2}. However building systems with many coupled qubits remains a challenge. Many demands must be met: reliable qubit storage, preparation and measurement, gate operations with high fidelities and low failure rate, scalability and accurate transportation or teleportation of states. The biggest problem is posed by decoherence which tends to destroy the desired quantum behaviour.

In recent years much has been done to address the problems posed by decoherence. An important step was the realisation that many systems possess a large subspace of decoherence-free states \cite{Palma,Zanardi97,Lidar98} which is well protected from the environment and provides ideal qubits. A great variety of approaches in manipulating the qubits within the decoherence-free subspace has been discussed in the literature. The optimal approach would be to employ only Hamiltonians leaving the decoherence-free subspace invariant and therefore not causing transitions into unwanted states \cite{Lidar00}. However they are in general hard to identify in physical systems. 

Alternatively, environment-induced measurements and the quantum Zeno effect \cite{Misra} can be used to avoid decoherence. The idea of {\em quantum computing using dissipation} \cite{beige4,letter,PachosWalther} employs the fact that the presence of spontaneous decay rates can indeed have the same effect as rapidly repeated measurements, whether the system is in a decoherence-free state or not, thus restricting the time evolution onto the decoherence-free subspace. As in linear optics quantum computing \cite{Knill,Franson}, the presence of measurements has the advantage that local operations on the qubits become sufficient for the implementation of universal quantum computation. This allows significant reduction in the experimental effort for the realisation of gate operations. For example, universal quantum gates between atomic qubits can be realised with the help of a single laser pulse \cite{cold}.

However, obtaining these advantages does not always require the presence of spontaneous decay rates in the system. In some cases, the presence of a strong interaction alone is sufficient to restrict the time evolution of a system onto a subspace of slowly-varying states \cite{vl,Agarwal}. An additionally applied interaction then causes an {\em adiabatic time evolution} inside this subspace. If the latter coincides with the decoherence-free subspace of the system, the effect of the strong interaction is effectively the same as the effect of continuous measurements whether the system is in a decoherence-free state or not \cite{Facchi}  and weak interactions can be used for the implementation of decoherence-free quantum gates. Concrete proposals for the implementation of this idea for ion-trap quantum computing and in atom-cavity systems can be found in Refs.~\cite{Beige,Marr}.

In this paper we discuss the positive role dissipation can play in a situation where quantum gates are implemented with the help of the adiabatic processes described above. Such a scheme is in general relatively robust against parameter fluctuations but any attempt to speed up operations results in the population of unwanted states. However, given the unwanted states possess spontaneous decay rates, the time evolution of the system can become a {\em dissipation-assisted adiabatic passage}. Even if operated rapidly, the system behaves as predicted for adiabatic processes. The reason is that the no-photon time evolution corrects for any errors due to non-adiabaticity \cite{Beige,Marr}. As a concrete example, we describe two-qubit gate operations in atom-cavity systems which can be performed twice as fast in the presence of certain decay rates without sacrificing their high fidelity and robustness. 

The atom-cavity systems provide a promising technology for quantum computing \cite{Walther}. The main sources of decoherence are dissipation of cavity photons with rate $\kappa$ and spontaneous decay from excited atomic levels with decay rate $\Gamma$. Some recent proposals for atom-cavity schemes attempt to minimise the population of excited states using strong detunings \cite{ZhengGuo,JanePlenioJonathan}; others use dissipation \cite{beige4,letter}. All these schemes are inherently slow which causes relatively high failure rates. Regarding success probabilities, the quantum computing schemes \cite{pellizzari,Shahriar,PachosWalther,Marr,You} perform much better. We believe that the dissipation-assisted adiabatic passages we describe in this paper contributes to the success of these schemes as well as being a key feature of the original proposal \cite{pellizzari}.

\section{Theoretical model} \label{sect:2}

Over the last three decades quantum optical experiments have been performed studying the statistics of photons emitted by laser-driven trapped atoms and effects have been found that would be averaged out in the statistics of photons emitted by a whole ensemble \cite{schmuss}. These experiments suggest that the effect of the environment on the state of the atoms is the same as the effect of rapidly repeated measurements and hence can result in a sudden change of the fluorescence of a single atom \cite{shelving}. From the assumption of measurements whether a photon has been emitted or not, the quantum jump approach \cite{heger} has been derived. This approach is equivalent to the Monte Carlo wave-function approach \cite{HeWi2} and the quantum trajectory approach \cite{HeWi3}.

\subsection{No-photon time evolution}

Suppose a measurement is performed on the free radiation field interacting with a quantum optical system initially prepared in $|\psi \rangle$. Under the condition of no photon emission and given that the free radiation field was initially prepared in its vaccum state $|0_{\rm ph} \rangle$, the (unnormalised) state of the system at $\Delta t$ equals, according to the quantum jump approach \cite{heger}, 
\begin{eqnarray} \label{def}
|0_{\rm ph} \rangle  \langle 0_{\rm ph} | \, U(\Delta t,0) \, |0_{\rm ph} \rangle |\psi \rangle 
&\equiv &  |0_{\rm ph} \rangle \, U_{\rm cond} (\Delta t,0) |\psi \rangle ~. \nonumber \\&&
\end{eqnarray}
The dynamics under the conditional time evolution operator $U_{\rm cond}(\Delta t,0)$, defined by this equation, can be summarised in a Hamiltonian $H_{\rm cond}$. This Hamiltonian is in general non-Hermitian and the norm of a state vector developing with $H_{\rm cond}$ decreases in time such that
\begin{equation} \label{P0}
P_0(t,\psi) = \| \, U_{\rm cond} (t,0) \, |\psi \rangle \, \|^2 
\end{equation}
is the probability for no emission in $(0,t)$. The non-Hermitian terms in the conditional Hamiltonian continuously damp away amplitudes of unstable states. This takes into account that the observation of no emission leads to a continuous gain of information about the state of the system. The longer no photons are emitted, the more unlikely it is that the system has population in excited states \cite{Cook,Hegerfeldt}. 

\subsection{Dissipation-assisted adiabatic passages} \label{xxx}

In the following we exploit the conditional no-photon time evolution to implement better gate operations. 
Quantum computing with non-Hermitian Hamiltonians can be performed with high success rates as long as the system remains to a good approximation in a decoherence-free subspace. A state $|\psi \rangle$ is decoherence-free if populating it cannot lead to a photon emission and  
\begin{equation} \label{p0cond}
P_0(t,\psi) \equiv 1
\end{equation}
for all times $t$ \cite{beige4}. The decoherence-free subspace is therefore spanned by all the eigenvectors of the conditional Hamiltonian $H_{\rm cond}$ with real eigenvalues and a time evolution with this Hamiltonian leaves the decoherence-free subspace invariant. 

In general it is very difficult to find a Hamiltonian that keeps the decoherence-free subspace invariant {\em and} can be used for the realisation of gate operations. However, it is always possible to add a weak interaction to the conditional Hamiltonian $H_{\rm cond}$, so generating a new conditional Hamiltonian $\tilde H_{\rm cond}$. 
As long as the additional interaction is weak, the decoherence-free subspace constitutes an invariant subspace of $\tilde H_{\rm cond}$ to a very good approximation. This can be exploited to generate an adiabatic time evolution inside the decoherence-free subspace according to the effective Hamiltonian 
\begin{equation} \label{heff}
H_{\rm eff} = I \!\!P_{\rm DFS} \, \tilde H_{\rm cond} \, I\!\!P_{\rm DFS} ~.
\end{equation}
This Hamiltonian can now be used for the realisation of quantum gates. A drawback of this idea is that the effective evolution, which happens on the time scale given by the weak interaction, is very slow. 

The essential idea of this paper is to ensure that the same net evolution within the decoherence-free subspace is realised with high fidelity even when the system is operated relatively fast, i.e.~outside the adiabatic regime, and despite the occurence of errors at intermediate stages. The form of Eq.~(\ref{heff}) assures that any error leads to the population of non decoherence-free states. As long as this population is small, there is a very high probability that it will be damped away during the no-photon time evolution. The system behaves effectively as predicted by adiabaticity and the underlying relatively fast process could be called a dissipation-assisted adiabatic passage \cite{Beige,Marr}.

Whenever a photon emission occurs, the computation fails and the experiment has to be repeated. Naively, one might expect that a finite probability for failure of the proposed scheme also implies a decrease of the fidelity of the gate operation. However, the fidelity under the condition of no photon emission remains close to unity for a wide range of experimental parameters \cite{cold}. Moreover, the non-Hermitian terms in $\tilde H_{\rm cond}$ inhibit transitions into unwanted states and stabilise the desired time evolution (\ref{heff}).

\subsection{Decoherence-free states with respect to cavity decay} \label{sect:2.3}

Let us now consider a concrete system. In the following, each qubit is obtained from two different ground states  $|0 \rangle$ and $|1 \rangle$ of the same atom. To implement two-qubit gate operations, the two corresponding atoms are moved inside the optical resonator where both see the same coupling constant $g$. Suppose, the 1-2 transition in each atom couples resonantly to the cavity mode and $b$ and $b^\dagger$ are the annihilation and creation operators for a single photon. The conditional Hamiltonian in the interaction picture with respect to the interaction-free Hamiltonian can then be written as 
\begin{eqnarray} \label{cond}
H _{\rm cond} &=& \hbar g \sum_{i=1}^2 b^\dagger |1\rangle _{ii} \langle2| + {\rm H.c.} \nonumber \\ 
&& -  {\textstyle {{\rm i} \over 2}} \hbar \Gamma \sum_i |2 \rangle_{ii} \langle 2| 
-  {\textstyle {{\rm i} \over 2}} \hbar \kappa b ^{\dagger} b ~.
\end{eqnarray}
The last two terms are the non-Hermitian terms of the Hamiltonian.

Using Eq.~(\ref{p0cond}) one can easily determine the de\-co\-he\-rence-free subspace of the atom-cavity system with respect to leakage of photons through the cavity mirrors \cite{beige4}. It is spanned by the eigenvectors of the conditional Hamiltonian $H_{\rm cond}$ with real eigenvalues assuming $\Gamma=0$ and includes all superpositions of the atomic ground states, i.e.~the qubits states of the system, and the maximally entangled atomic state 
\begin{equation} \label{a}
|a \rangle \equiv {\textstyle {1 \over \sqrt{2}}} \, (|12 \rangle - |21 \rangle) ~,~
\end{equation}
while the cavity is in its vacuum state $|0 \rangle_{\rm cav}$. Prepared in the state $|0 \rangle_{\rm cav} |a \rangle  \equiv |0;a\rangle$ the atoms do not interact with the cavity mode and therefore cannot transfer their excitation into the resonator. Thus no photon can leak out through the cavity mirrors. Populating only these states and performing the gate relatively fast, thereby reducing the possibility for spontaneous emission from the atoms with rate $\Gamma$, should result in relatively high gate success rates. 
  
\section{High-fidelity quantum computation in atom-cavity-systems} \label{sect:5}

\begin{figure}
\begin{minipage}{\columnwidth}
\begin{center}
\resizebox{\columnwidth}{!}{\rotatebox{0}{\includegraphics{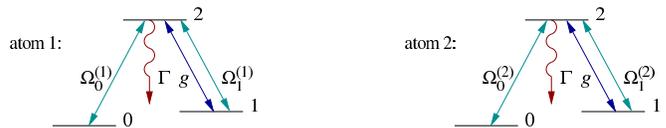}}}  
\end{center}
\vspace*{-0.5cm}
\caption{Level scheme of the two atoms inside the resonator. Each qubit is obtained from two different ground states $|0 \rangle$ and $|1 \rangle$ of one atom. To implement gate operations, laser fields with Rabi frequencies $\Omega_j^{(i)}$ can be applied exciting the $j$-2 transition of atom $i$.} \label{fig2}
\end{minipage}
\end{figure}

In this section we first neglect the spontaneous decay rates, assume $\kappa=\Gamma=0$ and aim at finding laser configurations and Rabi frequencies that result in a time evolution with the effective Hamiltonian (\ref{heff}) with respect to the decoherence-free subspace introduced in Section \ref{sect:2.3}. In the following, we consider the level configuration in Figure \ref{fig2} and denote the Rabi frequency with respect to the $j$-2 transition of atom $i$ as $\Omega_j^{(i)}$. Especially, we look for adiabatic processes where the two different time scales in the system are provided by the atom-cavity constant $g$ being a few orders of magnitude larger than the Rabi frequencies $\Omega_j^{(i)}$ of the applied laser fields,
\begin{equation}
\Omega_j^{(i)} \ll g ~.
\end{equation}
As concrete examples for gate implementations via dissipation-assisted adiabatic passages we describe possible realisations of a two-qubit phase gate and a SWAP operation. We then show that the same quantum gates can be operated twice as fast with decay rates, while maintaining fidelities above 0.98.

\subsection{Adiabatic elimination of unwanted states}

From the discussion in Section \ref{sect:2} we know already that the decoherence-free states of the system are the eigenstates of the conditional Hamiltonian (\ref{cond}) with real eigenvalues in the absence of the weak laser interaction. Consequently, they are also eigenstates of the interaction Hamiltonian of the system, namely
\begin{eqnarray} \label{nocond}
H &=& \hbar g \sum_{i=1}^2 b^\dagger |1\rangle _{ii} \langle2| + 
\sum_{i=1}^2 \sum_{j=0}^1 {\textstyle {1 \over 2}} \hbar \Omega^{(i)}_j  |j \rangle_{ii} \langle 2 | 
\nonumber \\
&& + {\rm H.c.} ~,~
\end{eqnarray}
to a very good approximation. It is therefore convenient to consider them in the following as basis states. To obtain a complete basis we introduce, in addition to Eq.~(\ref{a}), the symmetric state
\begin{equation}
|s \rangle \equiv {\textstyle {1 \over \sqrt{2}}} \, (|12 \rangle + |21 \rangle) ~.
\end{equation}
In the following, $|n;x \rangle$ denotes a state with $n$ photons in the cavity field and the atoms prepared in $|x \rangle$. 

We now write the state of the system as a superposition of the form
\begin{equation}
|\psi \rangle = \sum_{n,x} c_{n;x} \, |n;x \rangle 
\end{equation}
and first calculate the time evolution of the coefficients of the decoherence-free states $|0;00 \rangle$, $|0;01 \rangle$, $|0;10 \rangle$, $|0;11 \rangle$ and $|0;a \rangle$. The Hamiltonian (\ref{nocond}) yields
\begin{eqnarray} \label{slowest}
\dot c_{0;00} &=& - {\textstyle {{\rm i} \over 2}} \big[ \Omega_0^{(1)} c_{0;20} +  \Omega_0^{(2)} c_{0;02} \big]
~, ~ \nonumber \\  
\dot c_{0;01} &=& - {\textstyle {{\rm i} \over  2\sqrt{2}}} \big[ \Omega_0^{(1)} \big(c_{0;s} - c_{0;a} \big) 
+ \sqrt{2} \Omega_1^{(2)} c_{0;02} \big] ~,~ \nonumber \\ 
\dot c_{0;10} &=& - {\textstyle {{\rm i} \over  2\sqrt{2}}} \big[ \sqrt{2} \Omega_1^{(1)} \, c_{0;20} 
+  \Omega_0^{(2)} \big(c_{0;s} + c_{0;a} \big) \big] ~,~ \nonumber \\
\dot c_{0;11} &=& - {\textstyle {{\rm i} \over  2\sqrt{2}}} \big[  \big( \Omega_1^{(1)} + \Omega_1^{(2)} \big) c_{0;s} 
-  \big( \Omega_1^{(1)} - \Omega_1^{(2)} \big) c_{0;a} \big] ~,~ \nonumber \\  
\dot c_{0;a} &=&  - {\textstyle {{\rm i} \over  2\sqrt{2}}} \big[  \big( \Omega_1^{(1)} - \Omega_1^{(2)} \big) 
\big( c_{0;22} - c_{0;11} \big) - \Omega_0^{(1)} \, c_{0;01} \nonumber \\
&& + \Omega_0^{(2)} \, c_{0;10} \big] ~.
\end{eqnarray}
We furthermore consider the derivatives of the amplitudes of the states $|0;02 \rangle$, $|0;20 \rangle$, $|0;s \rangle$ and $|0;22 \rangle$,
\begin{eqnarray} \label{slow}
\dot c_{0;02} &=& - {\textstyle {{\rm i} \over 2}} \big[ \Omega_0^{(1)} c_{0;22} +  \Omega_0^{(2)} c_{0;00} +  \Omega_1^{(2)} c_{0;01} \big] - {\rm i} g c_{1;01} ~, ~ \nonumber \\  
\dot c_{0;20} &=& - {\textstyle {{\rm i} \over 2}} \big[ \Omega_0^{(1)} c_{0;00} +  \Omega_0^{(2)} c_{0;22} +  \Omega_1^{(1)} c_{0;10} \big] - {\rm i} g c_{1;10} ~,~ \nonumber \\ 
\dot c_{0;s} &=&  - {\textstyle {{\rm i} \over  2\sqrt{2}}} \big[  \big( \Omega_1^{(1)} + \Omega_1^{(2)} \big) 
\big( c_{0;11} + c_{0;22} \big) + \Omega_0^{(1)} c_{0;01} \nonumber \\
&& + \Omega_0^{(2)} c_{0;10} \big] - \sqrt{2} {\rm i} g c_{1;11} ~,~ \nonumber \\
\dot c_{0;22} &=& - {\textstyle {{\rm i} \over  2\sqrt{2}}} \big[ \big( \Omega_1^{(1)} + \Omega_1^{(2)} \big) c_{0;s} 
+  \big( \Omega_1^{(1)} - \Omega_1^{(2)} \big) c_{0;a} \nonumber \\
&& + \sqrt{2} \Omega_0^{(1)} c_{0;02} + \sqrt{2} \Omega_0^{(2)} c_{0;20} \big] - \sqrt{2} {\rm i} g c_{1;s} ~,~ 
\end{eqnarray}
and the states $|1;01 \rangle$, $|1;10 \rangle$, $|1;11 \rangle$, $|1;s \rangle$ and $|2;11 \rangle$,
\begin{eqnarray} \label{slower}
\dot c_{1;01} &=& - {\textstyle {{\rm i} \over 2\sqrt{2}}} \big[ \Omega_0^{(1)} \big( c_{1;s} - c_{1,a} \big)  
+ \sqrt{2} \Omega_1^{(2)} c_{1;02} \big] \nonumber \\
&& - {\rm i} g c_{0;02} ~, ~ \nonumber \\  
\dot c_{1;10} &=& - {\textstyle {{\rm i} \over 2 \sqrt{2}}} \big[ \sqrt{2} \Omega_1^{(1)} c_{1;20} +  \Omega_0^{(2)} \big( c_{1;s} +  c_{1,a} \big) \big] \nonumber \\
&& - {\rm i} g c_{0;20} ~,~ \nonumber \\ 
\dot c_{1;11} &=& - {\textstyle {{\rm i} \over  2\sqrt{2}}} \big[ \Omega_1^{(1)} \big( c_{1;s} - c_{1,a} \big)
+ \Omega_1^{(2)} \big( c_{1;s} + c_{1;a} \big) \big] \nonumber \\
&& - \sqrt{2} {\rm i} g c_{0;s} ~,~ \nonumber \\
\dot c_{1;s} &=& - {\textstyle {{\rm i} \over  2\sqrt{2}}} \big( \Omega_1^{(1)} + \Omega_1^{(2)} \big) 
\big( c_{1;11} +  c_{0;22} \big) \nonumber \\
&& - \sqrt{2} {\rm i} g \big( c_{0;22} + \sqrt{2} c_{2,11} \big) ~,~ \nonumber \\
\dot c_{2;11} &=& - {\textstyle {{\rm i} \over  2}} \big[ \Omega_1^{(1)} c_{2;21} +  \Omega_1^{(2)} c_{2;12} \big]   
- 2 {\rm i} g c_{1;s} ~.~
\end{eqnarray}
Suppose the system is initially prepared in a qubit state and the population of non-decoherence-free states remains of the order ${\cal O}(\Omega_j^{(i)}/g)$, we can easily eliminate all amplitudes that change on the fast time scale defined by the cavity coupling $g$. Setting their derivatives equal to zero and neglecting all terms of second order in $\Omega_j^{(i)}/g$ and smaller, the differential equations (\ref{slower}) yield
\begin{equation} \label{onkel}
c_{0;02} = c_{0;20} = c_{0;s} = c_{0;s} = 0 ~,~ c_{0;22} = - \sqrt{2} c_{2;11} ~.~ 
\end{equation}
Substituting this result into Eq.~(\ref{slowest}) and (\ref{slow}) we obtain further 
\begin{eqnarray} \label{longer}
&& \dot c_{0;00} = 0 ~,~ 
\dot c_{0;01} = {\textstyle {{\rm i} \over 2 \sqrt{2}}} \Omega_0^{(1)} c_{0;a} ~,~ \nonumber \\ 
&& \dot c_{0;10} = - {\textstyle {{\rm i} \over  2\sqrt{2}}} \Omega_0^{(2)} c_{1,a} ~,~ 
\dot c_{0;11} = {\textstyle {{\rm i} \over  2\sqrt{2}}} \big( \Omega_1^{(1)} - \Omega_1^{(2)} \big) c_{0;a} ~,~ \nonumber \\
&& \dot c_{0;a} = {\textstyle {{\rm i} \over  2}} \big[ \big( \Omega_1^{(1)} - \Omega_1^{(2)} \big) c_{0;11} 
+  \Omega_0^{(1)} c_{0;01} - \Omega_0^{(2)} c_{0;10} \big]  ~,~ \nonumber \\
&& \dot c_{0;22} = -{\textstyle {{\rm i} \over  2\sqrt{2}}} \big( \Omega_1^{(1)} - \Omega_1^{(2)} \big) c_{0;a} ~.~ 
\end{eqnarray}
This shows that the only way to restrict the time evolution of the system onto the decoherence-free subspace is to choose
\begin{equation} \label{schwester}
\Omega_1^{(1)} - \Omega_1^{(2)}=0 ~.
\end{equation}
Note that if this is not fulfilled, population leaks directly from the antisymmetric state $|0;a \rangle$ into the states $|0;22 \rangle$ and $|2;11 \rangle$. There is no mechanism in the system (see Eq.~(\ref{onkel})) that forbids the population of these states.

\begin{figure}
\begin{minipage}{\columnwidth}
\begin{center}
\resizebox{\columnwidth}{!}{\rotatebox{0}{\includegraphics{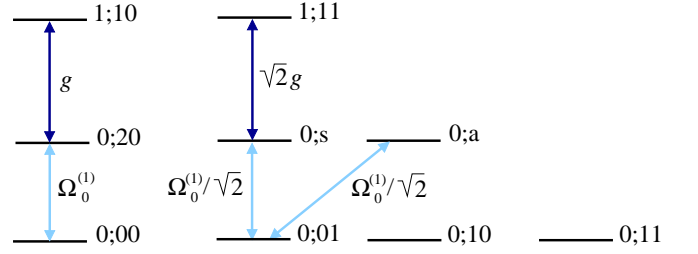}}}  
\end{center}
\vspace*{-0.5cm}
\caption{The relevant combined level scheme of the atom-cavity system for gate implementations with $\Omega_1^{(1)} = \Omega_1^{(1)} = 0$ contains only eleven states and the dynamics falls in several groups.} \label{daap1}
\end{minipage}
\end{figure}

The condition (\ref{schwester}) can easily be implemented with $\Omega_1^{(1)} = \Omega_1^{(2)}=0$ and the realisation of dissipation-assisted gate operations requires only a single laser field with the respective Rabi frequencies $\Omega_0^{(i)}$. As predicted before, the time evolution of the slowly-varying amplitudes of the system (\ref{longer}) can be summarised in the Hamiltonian 
\begin{eqnarray} \label{hopt}
H_{\rm eff} &=& 
- {\textstyle {1 \over 2 \sqrt{2}}} \, \big[ \hbar \Omega_0^{(1)} \, |01 \rangle \langle a|
- \hbar \Omega_0^{(2)} \, |10 \rangle \langle a| + {\rm H.c.} \big] ~,~ \nonumber \\&&
\end{eqnarray}
which coincides with the effective Hamiltonian given in Eq.~(\ref{heff}). The levels and transitions involved in the generation of the effective time evolution of the system are shown in Figure \ref{daap1}.

The corrections to the effective time evolution in first order ${\cal O}(\Omega_j^{(i)}/g)$ can be obtained by performing another adiabatic elimination of fast varying amplitudes using the differential equations (\ref{slow}) and setting $\dot c_{0;02} = \dot c_{0;20} = \dot c_{0;s} = 0$. During gate operations, a small population, given by
\begin{eqnarray} \label{colder}
&& c_{1;01} = - {\Omega_0^{(2)} \over 2g} \, c_{0;00} ~,~ 
c_{1;10} = - {\Omega_0^{(1)} \over 2g} \, c_{0;00} ~,~ \nonumber \\
&& c_{1;11} =  - {\Omega_0^{(1)} \over 4g} c_{0;01} - {\Omega_0^{(2)} \over 4g} c_{0;10} ~,~
\end{eqnarray}
accumulates in the states $|1;01 \rangle$, $|1;10 \rangle$ and $|1;11\rangle$. There are {\em two} ways to keep these errors small and we use both of them in the following. First, one should turn the laser field off slowly such that ${{\rm d} \over {\rm d}t} \Omega_0^{(i)} (T)=0$ with $T$ being the gate operation time. Then the system can adapt to the changing parameters and the unwanted amplitudes (\ref{colder}) vanish at time $T$. Second, the presence of a finite cavity decay rate $\kappa$ can be used to damp away photon population in the cavity mode during the no-photon time evolution. However, $\kappa$ should not be too large in order not to disturb the adiabatic evolution.

We should also mention that there is another way to guarantee that the system remains within the decoherence-free subspace. This requires the cavity decay rate $\kappa$ to be of the same order as $g$, which suppresses the population of {\em all} non decoherence-free states (including the states $|0;22 \rangle$ and $|2;11 \rangle$), independent of the choice of the Rabi frequencies $\Omega_j^{(i)}$ \cite{letter}. Indeed, it has been shown that quantum computing using dissipation leads to a realm of new possibilities for the implementations of quantum gate operations \cite{letter}. However, the approach we consider here yields schemes with much shorter gate operation times than the ones reported in Ref.~\cite{letter}. 

\subsection{The realisation of two-qubit gates}

A simple gate operation that can easily be implemented with the effective Hamiltonian (\ref{hopt}) is the quantum phase gate with
\begin{equation} \label{phase}
U_{\rm phase} = |00 \rangle \langle 00| - |01 \rangle \langle 01| + |10 \rangle \langle 10| + |11 \rangle \langle 11| ~.~ 
\end{equation}
This operation changes the state $|1\rangle$ of the second atom into $-|1\rangle$ provided that the first atom is in $|0 \rangle$. Suppose there is only one laser field coupling to the 0-2 transition of atom 1 such that $\Omega_0^{(2)} = 0$ and
\begin{equation} \label{nocond2}
H_{\rm eff} = - {\textstyle {1 \over 2 \sqrt{2}}} \, \hbar \Omega_0^{(1)} \big[  |01 \rangle \langle a| + {\rm H.c.}  \big] ~.~
\end{equation}
The implementation of the unitary operation (\ref{phase}) then only requires
\begin{equation} \label{opa}
\int_0^T \Omega_0^{(1)} {\rm d}t = 2\sqrt{2}\pi 
\end{equation}
as one can see from solving the time evolution of the Hamiltonian (\ref{nocond2}). In the following we choose
\begin{equation} 
\Omega_0^{(1)}(t) = 2\Omega_{\rm max} \, \sin^2 \left( {\textstyle {1 \over 2 \sqrt{2}}} \Omega_{\rm max} t \right) ~,~
\end{equation}
and  $T= 2\sqrt{2} \pi/\Omega_{\rm max}$.

\begin{figure}
\begin{minipage}{\columnwidth}
\begin{center}
\resizebox{\columnwidth}{!}{\rotatebox{0}{\includegraphics{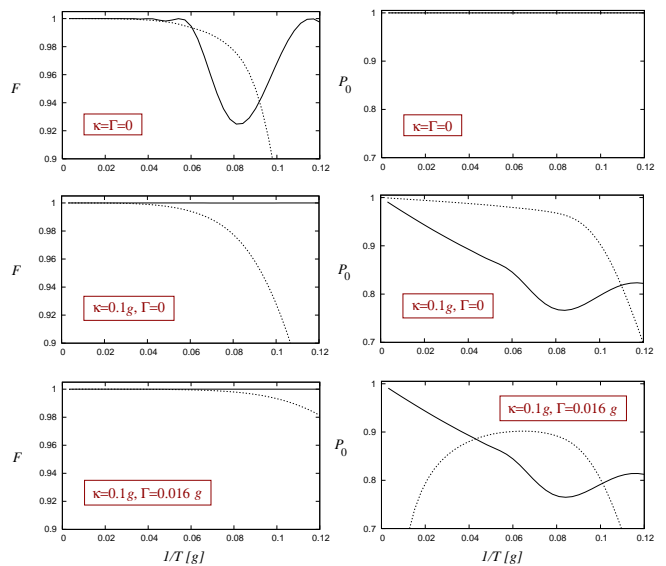}}}  
\end{center}
\vspace*{-0.5cm}
\caption{Fidelity $F$ and success rate $P_0$ of the phase gate (\ref{phase}) as a function of the inverse gate operation time $1/T$ for different spontaneous decay rates $\kappa$ and $\Gamma=0$ and for the initial qubit states $|00\rangle$ (solid line) and $|01 \rangle$ (dashed line).} \label{daap3}
\end{minipage}
\end{figure}

Given $\Omega_0^{(2)}=0$, the amplitudes of the states $|0;10 \rangle$ and $|0;11 \rangle$ remain unchanged while a minus phase is added to the state $|0;01 \rangle$ after undergoing an adiabatic transition to the excited state $|0;a \rangle$ and back (see Figure \ref{daap1}). That this is indeed the case can be seen in Figure \ref{daap3}, which results from a numerical solution of the no-photon time evolution of the system with the Hamiltonian 
\begin{eqnarray} \label{condi}
\tilde H _{\rm cond} &=& \sum_{i=1}^2 \hbar g b^\dagger |1\rangle _{ii} \langle2| 
+ {\textstyle {1 \over 2}} \hbar \Omega^{(1)}_0  |0 \rangle_{11} \langle 2 | + {\rm H.c.} \nonumber \\ 
&& - \sum_{i=1}^2  {\textstyle {{\rm i} \over 2}} \hbar \Gamma |2 \rangle_{ii} \langle 2| 
-  {\textstyle {{\rm i} \over 2}} \hbar \kappa b ^{\dagger} b 
\end{eqnarray}
for the initial qubit states $|00 \rangle$ and $|01 \rangle$. For $\kappa=\Gamma=0$, fidelities above $0.99$ are achieved for gate operation times  $1/T < 0.05 \, g$. 

Furthermore, the phase gate can be operated in the presence of decay rates. The presence of decay rates allows the scheme to be run twice as fast while maintaining the same high fidelity. Looking at the initial state $|0;00\rangle$, it can be seen that with $\kappa=0$ the fidelity dips to just over $92 \, \%$ but with $\kappa = 0.1 \, g$ the fidelity is always one. The reason is that the presence of cavity decay rates damps away all the population in the states $|0;20 \rangle$ and $|1;10 \rangle$, at the latest at the end of the operation. Maintaining a fidelity close to $1$ comes at cost of a significantly lowered success probability, which can be even lower than $80 \, \%$. For larger values of $\kappa$ the success probability would be even smaller.

The fidelity given the initial state $|0;01\rangle$ is only marginally helped by the non-zero $\kappa$. However, a small $\Gamma = 0.016 \, g$ can further reduce the errors. The reason is that in the non-adiabatic regime unwanted population accumulates in the excited atomic state $|a\rangle$ which is then taken care of by $\Gamma >0$. The competition between greater speed tending to reduce the probability of a failure due to $\Gamma$ and an increased error due to non-adiabaticity leads to a maximum for the gate success rate of $P_0 = 90 \, \%$ at $1/T=0.065 \, g$.

Another quantum gate that can easily be implemented with the effective Hamiltonian (\ref{hopt}) is the SWAP operation with
\begin{equation} \label{SWAP}
U_{\rm SWAP} = |00 \rangle \langle 00| + |10 \rangle \langle 01| + |01 \rangle \langle 10| + |11 \rangle \langle 11| ~.~ 
\end{equation}
Unlike the controlled phase gate, the SWAP gate is not universal. Nevertheless, this operation can be very useful since it exchanges the states of two qubits without that the corresponding atoms have to physically swap their places. To implement the time evolution (\ref{SWAP}) one should choose $\Omega_0^{(1)} = \Omega_0^{(2)}$ and individual laser addressing of the atoms is not required. The effective Hamiltonian becomes in this case 
\begin{equation} \label{nocond3}
H_{\rm eff} = - {\textstyle {1 \over 2 \sqrt{2}}} \, \hbar \Omega_0^{(1)} \big[ |01 \rangle \langle a| +  |10 \rangle \langle a| + {\rm H.c.}  \big] 
\end{equation}
and implements a SWAP operation if
\begin{equation} \label{oma}
\int_0^T \Omega_0^{(1)} {\rm d}t = 2 \pi ~.
\end{equation}
In the following we choose
\begin{equation} 
\Omega_0^{(1)}(t) = 2\Omega_{\rm max} \, \sin^2 \left( {\textstyle {1 \over 2}} \Omega_{\rm max} t \right) ~,~
\end{equation}
and $T= 2 \pi/\Omega_{\rm max}$. 

\begin{figure}
\begin{minipage}{\columnwidth}
\begin{center}
\resizebox{\columnwidth}{!}{\rotatebox{0}{\includegraphics{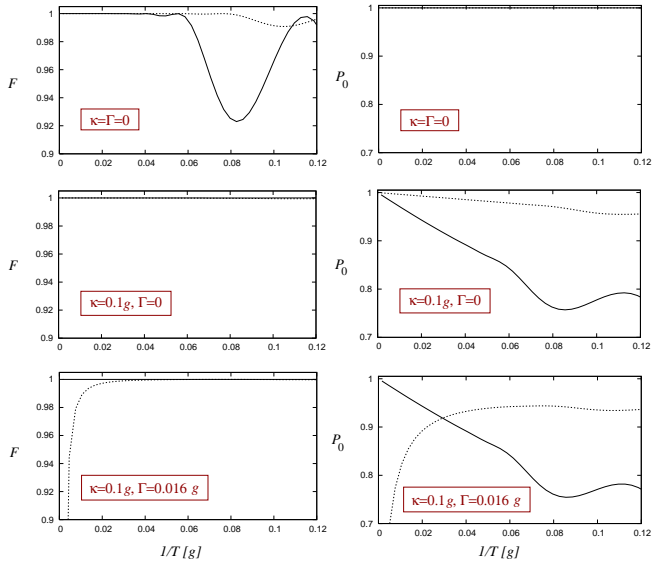}}}  
\end{center}
\vspace*{-0.5cm}
\caption{Fidelity $F$ and success rate $P_0$ of a single SWAP operation as a function of the inverse gate operation time $1/T$ for the initial qubit states $|00\rangle$ (solid line) and $|01 \rangle$ (dashed line).} \label{fig4}
\end{minipage}
\end{figure}

Figure \ref{fig4} shows the results of a numerical solution of the no-photon time evolution of the system with the Hamiltonian 
\begin{eqnarray} \label{condi2}
\tilde H _{\rm cond} &=& \sum_{i=1}^2 \hbar g b^\dagger |1\rangle _{ii} \langle2| 
+ {\textstyle {1 \over 2}} \hbar \Omega^{(i)}_0  |0 \rangle_{ii} \langle 2 | + {\rm H.c.} \nonumber \\ 
&& - \sum_{i=1}^2  {\textstyle {{\rm i} \over 2}} \hbar \Gamma |2 \rangle_{ii} \langle 2| 
-  {\textstyle {{\rm i} \over 2}} \hbar \kappa b ^{\dagger} b
\end{eqnarray} 
for the same parameters as in Figure \ref{daap3}. For $\kappa=\Gamma=0$, fidelities above $0.99$ are achieved as long as $1/T < 0.06 \, g$. In the presence of the decay rates $\kappa = 0.1 \, g$, $\Gamma = 0.016 \, \Gamma$ and for $1/T=0.04 \, g$, the fidelity is well above $99 \, \%$ while $P_0 > 90 \, \%$. The results for the SWAP operation are very similar to the results for the phase gate. The reason for this is that the relevant combined level scheme of the system is in both cases about the same (see Figure \ref{daap1}).

\section{Conclusions}

This paper discusses how dissipation can help to increase the performance of quantum gate operations. The focus is on atoms with a $\Lambda$-type level configuration trapped in an optical cavity but the conclusions can be applied more generally. The concrete examples considered here are two-qubit operations including a controlled phase gate and the SWAP operation. For $g^2 = 625 \, \kappa \Gamma$, we showed that the gate success rate can be nearly as high as $90 \, \%$ while the fidelity is well above $99 \, \%$. Improved results have only been obtained in Refs.~\cite{PachosWalther,You,pellizzari}. The main problem of the proposed quantum computing scheme is spontaneous emission from the atoms. Nevertheless, it is very simple, as it requires only a single laser field, and it is widely robust against parameter fluctuations (see Eqs.~(\ref{opa}) and (\ref{oma})).

In the absence of dissipation, the quantum gates described employ adiabaticity arising from the different time scales set by the laser Rabi frequencies and atom-cavity coupling constant. However, in the presence of decay rates, like $\kappa = 0.1 \, g$ and $\Gamma = 0.016 \, g$, the scheme can be operated about twice as fast. The reason is that the underlying process becomes a dissipation-assisted adiabatic passage \cite{Beige,Marr}. Even operated outside the adiabatic regime, the no-photon time evolution corrects for errors and the system behaves as predicted by adiabaticity. The high fidelity comes at a cost of a finite gate failure rate. However, as long as one can detect whether an error occured or not and repeat the computation whenever necessary, this approach can be used to implement high-fidelity quantum computing even in the presence of dissipation.

\begin{acknowledgements}  
A.B. thanks the Royal Society and the GCHQ for funding as a James Ellis University Research Fellow. This work was supported in part by the UK Engineering and Physical Sciences Research Council and the European Union through QGATES and CONQUEST.
\end{acknowledgements}

\end{document}